# Inverse Energy Cascade in Turbulent Taylor-Couette Flows

Changquan Zhou[1,2], Hua-Shu Dou[1]*, Lin Niu[1], Wenqian Xu [1,3]

[1]Faculty of Mechanical Engineering, Zhejiang Sci-Tech University, Hangzhou, Zhejiang, 310018, China
[2]School of Mechanical Engineering, Zhejiang University of Water Resources and Electric Power,
Hangzhou, Zhejiang, 310018, China
[3]School of Mechanical Engineering, Hangzhou Dianzi University, Hangzhou, Zhejiang 310018, China
*Email: huashudou@zstu.edu.cn

**Abstract** The inverse energy cascade in turbulent Taylor-Couette flow is studied in line with the results of the large eddy simulation. The simulation results show that the inverse energy cascade first occurs within the core region of the flow channel of the Taylor-Couette flow at higher Reynolds number. It is uncovered that this phenomenon is induced by the pulsed zero shear stress resulting from the singularities of the Navier-Stokes equation. In the core area between the two cylinders, the shear stress is nearly zero at higher Reynolds number. The turbulence generated there has high turbulent energy due to discontinuity of the tangential velocity. Since the energy transfer between the fluid layers is inhibited due to the low shear stress, the turbulent energy cannot be transferred along the radial direction, and small-scale vortices with high turbulent energy are produced. These small-scale vortices are located with the large-scale vortices and cannot be dissipated owing to low shear stress. A peak in the energy spectrum at middle frequency (or wave number) is formed due to the concentration of the small-scale vortices. As the number of the singular points of the Navier-Stokes equation increases with the increasing Reynolds number, the region with zero shear stress expands along the radial direction, intensifying nonlinear instability and energy accumulation. This, in turn, leads to more prominent peaks in the energy spectrum, resulting in a more pronounced inverse energy cascade.

## 1.Introduction

The mechanism of turbulence is still not fully understood despite over 100 years of researches. The energy cascade is a fundamental mechanism describing the exchange of energy between different scales of motions in turbulence [1]. It describes how energy is gradually transferred from large-scale vortices to small-scale vortices through nonlinear interactions in turbulent flow, and finally converted into thermal energy at the dissipation scale. This is crucial for grasping the intrinsic nature of turbulence. Richardson's poem provides a vivid metaphor for the energy cascade [2]: "Big whorls have little whorls, which feed on their velocity; and little whorls have lesser whorls, and so on to viscosity." Kolmogorov obtained the renowned the power law of -5/3 of the energy spectrum in inertial subrange based on the statistical theory of homogeneous isotropic turbulence, known as the K41 theory [3]. In the energy cascade, large-scale vortices transfer energy to smaller-scale vortices through inertial effects, forming a series of continuous scale chains. This transfer is manifested in the spectral space as a gradual decrease in the energy spectrum, that is, the energy density decreases with the increase of the wave number. According to the K41 theory, the slope of the turbulent energy versus the wave number should be in a power law of -5/3 in the subrange, which provides a theoretical basis for understanding the energy cascade.

However, it is found that this classical theory does not universally apply to all the cases of turbulence. Kraichnan [4] was the first to theorize the existence of an inverse energy cascade in two-dimensional turbulent flows, attributing this phenomenon to the conservation of enstrophy in middle scales. It was demonstrated that the direct cascade of energy is blocked due to the conservation of enstrophy, causing the energy to be transported from small scales to large scales through an inverse cascade. This theoretical result has been confirmed by numerous experiments and numerical simulations [5-11]. Chen et al. [12] and Xiao et al. [13] employed direct numerical simulation techniques, in conjunction with filtered spatial decomposition of velocity fields and local energy fluxes, to uncover the underlying physics of the inverse cascade. Their research corroborated Kraichnan's proposition that vortex thinning is a pivotal mechanism for the inverse energy cascade, offering novel insights into the intricacies of turbulent energy transfer.

In the three-dimensional fluid dynamics, the kinetic energy and the helicity, as the conserved quantities of inviscid flow, have incited significant research interest due to their roles in the energy transfer mechanism. The influence of helicity on the energy cascade, the pathway through which the energy is conveyed from larger to smaller scales, is particularly pronounced. An increase in helicity effectively decelerates the energy transfer from large to small scales, consequently diminishing the rate of energy dissipation. This phenomenon is rooted in the pivotal role of nonlinear interactions within the energy transfer. Kraichnan and Panda [14] have robustly substantiated through comprehensive numerical simulations of isotropic turbulence that nonlinear effects can indeed exert a suppressive influence on turbulence. Chen et al. [15] and Chen et al. [16] have utilized direct numerical simulation (DNS) to investigate the statistical characteristics of energy and helicity fluxes in three-dimensional turbulent flows. Their work has uncovered the intermittent nature of helicity fluxes and the variations in helicity modes as energy cascades of large scales to smaller scales. Hu et al. [17], by integrating helicity into rotating turbulence, first identified that helicity significantly inhibits the inverse energy cascade induced by rotation.

Regarding the generation mechanisms of the inverse energy cascade in three-dimensional turbulent flows, various interpretations exist in literature. Some researchers attribute it to the vortex structures in the flow, the modes of energy input, and the flow boundary conditions. Cheikh et al. [18] explored the dynamics of small-scale energy cascade in homogeneous isotropic turbulence (HIT) through numerical simulations, finding that the rotation and translational motion of small-scale structures are highly dependent on the energy cascade. Wit et al. [19] carried out both theoretical studies and numerical simulations, and proposed that non-dissipative mechanisms (such as odd viscosity) could terminate turbulent cascades. This mechanism accumulates energy at intermediate scales and leads to the emergence of structures with specific characteristic sizes. Galtier [20] found that when a three-dimensional magnethydrodynamic (MHD) flow is excited at an intermediate scale $k_f$, it can cause an inverse cascade of magnetic helicity following a $k^{-2}$ spectrum, as well as a direct cascade of energy. However, for such a dual cascade to take place, the excitation must encompass both the injection of magnetic helicity and energy fluxes. Biferale et al. [21] discussed the phenomenon of the inverse energy cascade in three-dimensional isotropic turbulence and observed that under certain conditions, energy can migrate from smaller scales to larger ones, which is closely tied to boundary conditions. The studies of Xia et al. [22] and Słomka and Dunkel [23] indicate that vortex structures in three-dimensional turbulence significantly influence the phenomenon of inverse energy cascade. When the flow exhibits strong rotation or shear, it may facilitate the coalescence and expansion of small-scale vortices, thus inducing the inverse energy cascade phenomenon.

The energy spectrum in the Taylor-Couette flow has been studied extensively by several researches. Lewis and Swinney [24] observed through experiments that, in contrast to fully developed isotropic turbulence and turbulent flows within pipes or channels, Taylor-Couette flow lacks an inertial range, and the energy spectrum does not display the -5/3 power law of Kolmogorov even at high Reynolds numbers, suspecting that rotation inhibits the energy cascade in Taylor-Couette flow. Shtilman et al. [25] first experimentally confirmed the phenomenon of inverse energy cascade in Taylor-Couette flow. They demonstrated in a rotating fluid when the rotational velocity is sufficiently high, the fluid motion takes on quasi-two-dimensional characteristics. This two-dimensionalization prevents vortex stretching, thereby leading to the occurrence of inverse energy cascade. Dong [26] performed direct numerical simulation on Taylor-Couette flow with a radius ratio of η=0.5, the inner cylinder rotating at Re=8000 and the outer cylinder stationary. A small-scale peak in the spatial energy spectrum of the radial velocity is identified, where the wave number k at the peak corresponds to the spacing of Taylor vortices, yet without explaining the cause of this small-scale peak. Van Hout and Katz [27], through experiments with counter-rotation, also failed to identify an inertial range with a -5/3 slope on the energy spectrum, attributing this to the physical size constraints of large-scale vortices imposed by the experimental setup. Huisman et al. [28] experimentally found that for the case of a rotating inner cylinder and a stationary outer cylinder, no inertial range with a slope of −5/3 was observed in the energy spectra. For cases with rotation ratios

μ=-0.6, -0.8, and -1, a −5/3 power law could be observed, but they did not investigate the peaks appearing in the spectral range for various scales. The experimentally results at high Reynolds numbers in Froitzheim et al. [29] showed that small-scale peak exists on the turbulent energy spectra that is located at wavenumbers around 10~20 independent of the radial coordinate (Fig.15 in [29]). However, its amplitude strongly varies with positions within the Taylor vortex rolls.

The generation and evolution of the inverse energy cascade phenomenon are influenced by a multitude of factors including rotational effects, geometric constraints, and boundary conditions. These effects are intricately intertwined, collectively shaping the characteristics of the inverse cascade. Nonetheless, the profound relationship between the inverse cascade phenomenon and the the intrinsic singularities of fluid dynamics governing equations remains incompletely understood. To gain a profound understanding of the inverse cascade phenomena, it is important to consider not only accounting for the collective impact of the aforementioned physical factors but also to study the inherent nonlinear dynamics of the fluid dynamics equations. By comprehending the physical mechanisms underlying the inverse cascade phenomena, efficient control strategies can be designed in engineering, such as drag reduction, enhanced heat transfer, and improved fluid mixing efficiency, etc.

In this study, the mechanism of the inverse energy cascade occurred in Taylor-Couette flows will be studied from the aspects of turbulent generation, in line with the numerical simulations with large eddy simulation (LES). A new angle of view on the problem opened a solution to the inverse energy cascade. It is found that there is no transfer of turbulent energy from small vortices to large vortices in the inverse energy cascade, but the accumulation of high energy small vortices within large vortices at special flow conditions.

## 2 Theory revisiting on singularity generation in Taylor-Couette flow
### 2.1 Work done between fluid layers in Taylor-Couette flow and singularity

In this study, the outer cylinder is kept static and the inner cylinder is rotating with an angular velocity $\omega_1$, the radius ratio between the two cylinders is $\eta = R_1/R_2$, where $R_1$ is the radius of the inner cylinder and $R_2$ is the radius of the outer cylinder. The components of the velocity in the tangential, radial, and axial directionsare expressed by $u, v$, and $w$, respectively. For the base flow, assuming the flow is $\frac{\partial}{\partial \theta} = 0$, $v=0$, the flow is laminar, and there is no axial flow for the initial laminar flow.

Dou [30-32] found that the mechanism of turbulent transition in parallel pressure-driven flow (channel) is resulted from the singularity at the location of zero energy transfer between fluid particles. At the singularity, discontinuity of streamwise velocity occurs. This type of singularity is the place where the streamwise velocity drops to zero suddenly rather than blowing up to tending to infinite [30-32]. In shear-driven flows, such as the plane Couette flow and the Taylor-Couette flow, similar principle can be applied, that if there is no wok done instantly on the fluid particle, the velocity of the fluid particle will suddenly become zero in temporal evolution.

For the circumferential flow in the Taylor-Couette flow, the flow in the gap of cylinders can be treated as a layer by layer from the inner cylinder to the outer cylinder. For steady flow, the work done in per unit length by the shear stress along the streamwise direction by one layer to the neighboring layer was obtained in [32-34] for general cases of shear-driven circular flows, which is equal to the rate of the mechanical energy loss in circumferential flow,

$$\frac{\partial W}{\partial s} \equiv \frac{\tau}{u}\frac{\partial u}{\partial r} - \frac{\tau}{r} = \frac{\tau}{u}(\frac{\partial u}{\partial r} - \frac{u}{r}) = \frac{\phi}{u}, \qquad (1)$$

where $W$ is the work done in per unit length by the shear stress along the circumferential direction, s refers to the streamwsie direction and here it is the tangential direction, $\tau$ is the shear stress, and $\phi$ is the dissipation rate.

In [32-34], considering the steady flow in the gap of the Taylor-Couette flow, the net work done by the shear stress along the circumferential direction, should be equal to the energy loss of the fluid element layer. Now, for the case of unsteady flow, Eq.(1) should be hold too at the time moment with $\frac{\partial u}{\partial t}=0$ within a period of disturbance.

The shear stress and the dissipation rate are expressed as follow, respectively,

$$\tau = \mu\left(\frac{\partial u}{\partial r} - \frac{u}{r}\right), \qquad (2)$$

$$\phi = \mu\left(\frac{\partial u}{\partial r} - \frac{u}{r}\right)^2 \qquad (3)$$

where $\mu$ is the dynamic viscosity of fluid.

Equation (1) is applicable to flows for one cylinder rotating and the other at rest as well as the case for two cylinders rotating in opposite directions.

In the gap between the two cylinders, the fluid flow is due to the work done of shear stress by the neighbor layer in shear-driven flow [32, 33]. According to the principle of energy conservation, if the work done is $\partial W/\partial s = 0$ at a location, there is no energy transfer between fluid layers at this position in the shear-driven flow. Thus, this layer of fluid will stop and the velocity at this position will immediately become zero theoretically in temporal evolution.

If there is no work transfer at a position in the flow, $\partial W/\partial s = 0$, it is obtained from Eqs. (1) to (3) that,

$$\phi = 0, \ \tau = 0, \qquad (4)$$

Combining Eq.(4) with Eqs.(2) and (3), we obtain at this position,

$$\frac{\partial u}{\partial r} = \frac{u}{r}, \text{ and } u = r\frac{\partial u}{\partial r} \qquad (5)$$

It is seen from Eq.(5) that $\partial u/\partial r$ is not zero at the said position if u is not zero within the gap. Conversely, if $\partial u/\partial r \neq 0$ at this location, the velocity will be no-zero in terms of Eq.(5). However, according to the principle of energy conservation (Eq.(1)), in the temporal evolution, the circumferential velocity will immediately become zero theoretically at next moment after $\tau=0$,

$$u = 0 \tag{6}$$

The interval of time for $u=0$ is zero theoretically.

In such away, discontinuity of the circumferential velocity is produced along the radial direction at the moment. Owing to the effect of the viscosity, the circumferential velocity $u$ may not be completely zero, but sharp negative spike is produced, which is similar to that in plane Poiseuille flow [30, 31]. Because the velocity is not differentiable at the spike, this position forms a singularity in the flow field. Therefore, the position with $\tau=0$ in temporal evolution in the flow field is a singular point of the Navier-Stokes equation.

In summary, the singularity in the gap of the Taylor-Couette flow occurs at a position satisfying the following conditions, (a) the temporal term $\frac{\partial u}{\partial t}=0$, i.e., at the top or bottom of the wave form of the velocity; (b) the shear stress $\tau=0$. At such positions, the tangential velocity is theoretically zero u=0.

Dou et al derived an energy gradient function to express the feature of flow stability for shear-driven flows [32-34],

$$K = \frac{\partial E/\partial n}{\partial W/\partial s} = \frac{\partial E/\partial n}{\phi/u} \tag{7}$$

where $E = p + 0.5\rho V^2$ is the total mechanical energy, $p$ is the static pressure, and $V$ is the total velocity. At the position of $\partial W/\partial s = 0$, the value of $K$ is infinite. Thus, the said position is the singular point of the energy gradient function.

For the case of only the inner cylinder rotating, the Reynolds number based on the gap width is defined as follow,

$$\text{Re}_h = \frac{u_{\theta 1} h}{\nu}, \tag{8}$$

where, $h = R_2 - R_1$ is the gap width, and $R_1$ and $R_2$ are the radii of the inner and outer cylinders, respectively, $u_{\theta 1}$ is the tangential velocity of the inner cylinder, and $\nu$ is the kinematic viscosity.

### 2.2 Energy transfer in Taylor-Couette flow

Model for fluid flow in Taylor-Couette device is presented in **Figure 1**. In incompressible flow, the flow field can be expressed by large quantity of fluid elements. The fluid flow of each element is driven by a "driving energy" which is acted on each element, separately. If this "driving energy" suddenly changes to zero, the motion of the fluid element stops and the corresponding velocity becomes zero theoretically. In this paper, the "driving energy" acting on each element is named as "pseudo-force."

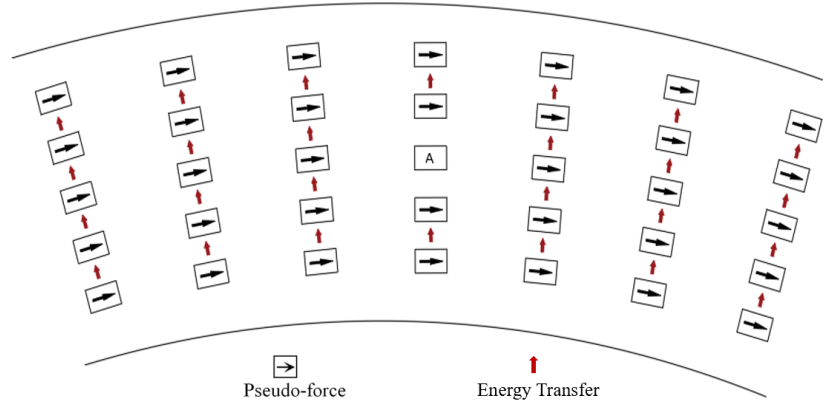

**Figure 1** Schematic of "energy transfer" and "pseudo force" acting on fluid element in Taylor-Couette flow.

For Taylor-Couette flow, take an element in **Figure 1**, the energy difference across the length $\Delta x$ along the streamwise direction is

$$\frac{\partial W}{\partial x}\Delta x = \Delta W$$

The area of the cross section of the element is $\Delta A = \Delta y \Delta z$. The energy consumed on the element with length $\Delta x$ in x direction is

$$\Delta F = \Delta W \cdot \Delta A = \frac{\partial W}{\partial x}\Delta x \Delta y \Delta z = \frac{\partial W}{\partial x}\Delta V \qquad (9)$$

where $\Delta V = \Delta x \Delta y \Delta z$ is the volume of the element. In Eq.(9), if $\Delta F$ is zero, the velocity u is theoretically zero according to the discussion on Eq.(1).

In Eq. (9), since $\Delta F$ has the unit of force (unit: N), it is called the "pseudo-force" force acting on the element in this paper. It may be easily understood that if the "pseudo-force" acting on an element is zero, the element will stop to move and the velocity is tending to zero.

As is well known, for viscous fluid flow, fluid particle consumes energy with the flow forward due to viscous friction. For Taylor-Couette flow, to sustain the fluid flow, energy must be supplied from external for each element. Along a streamline direction (x expresses the streamline direction and *u* expresses the velocity), the magnitude of the velocity depends on the work done by external shear stresses (**Figure 1**).

According the energy gradient theory, the transition of a laminar flow to turbulence must undergo a singularity of the Navier-Stokes equations. At this singular point, the energy gradient function K reaches infinite, and the energy transfer between fluid layers at this point vanishes. The instability in shear flow is dominated by the energy gradient function K [31, 32, 34, 35]. Since the value of K reaches its maximum on the moving wall in the basic laminar flow, the instability of the laminar flow may be initiated near the wall. However, the initiation of transition from laminar flow to turbulence is not necessarily at the same position as that for instability initiation of laminar flow, but it begins at the central zone between the walls according to the singularity criterion.

Actually, the occurrence of instability of a laminar flow is not equal to the starting of transition of the laminar flow to turbulence. After the instability sets in for a laminar flow, the flow may become another laminar flow, or may become turbulence, depending on that the criterion of transition is satisfied or not.

The distributions of the time-averaged tangential velocity and the time-averaged shear stress at the critical condition for spike generation are shown in Figure 2. In Fig.2(a) and (b), the averaged velocity and the averaged shear stress at the critical condition are shown, respectively. In the temporal evolution, both the tangential velocity and the shear stress will vary as a wave form. At the critical condition to producing singularity, both of them will become zero theoretically. Due to the fluid viscosity, spikes will be generated actually. Figure 2 (c) shows schematically the temporal variation of instantaneous tangential velocity at the critical condition. When the shear stress is instantaneously zero in the core region of the flow channel, the tangential velocity at the position with the zero shear stress will become zero theoretically, and finally a negative spike is produced. In such away, the velocity discontinuity is generated in temporal evolution.

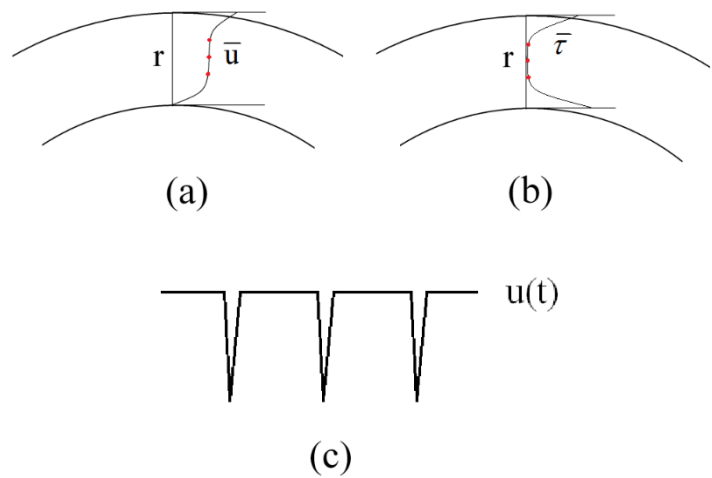

**Figure 2** Schematic of distributions of velocity and shear stress at the critical condition for spike generation in Taylor-Couette flow. (a) Averaged tangential velocity; (b) Averaged shear stress; (c) Temporal variation of instantaneous tangential velocity at critical condition at a location as shown by the red points in the figure (a) and (b). At these locations, negative spikes of streamwise velocity are produced.

When a discontinuity (spike) is produced on the velocity profile, the transfer of the mechanical energy (or angular momentum) between fluid layers is discontinued. In such away, the transfer of the angular momentum in rotating Taylor-Couette flow is reduced in transitional flow. If the Reynolds number is sufficiently high, the flow in the gap between the two cylinders may transit to turbulence with the spike as a source of instability. In fully developed turbulence, there are numerous singularities within the gap, and the transfer of the angular momentum is much constrained.

The relation of the total mechanical energy and the angular momentum in Taylor-Couette flow can be written below,

$$\frac{\partial (ru)^2}{\partial r} = \frac{2}{\rho} r^2 \frac{\partial E}{\partial r} \tag{10}$$

where $E = p + 0.5\rho V^2$ is the total mechanical energy per unit volume. In the gap between the cylinders, if the velocity profile becomes flat at a position such that the work done between fluid layers is zero, the shear stress will be near zero. Then, the streamwise velocity will be zero immediately at the next moment. In such away, discontinuity of tangential velocity occurs and negative spike is produced.

It is remembered that the total mechanical energy $E$ represents the energy of incompressible fluid in straight flow, while the square of angular momentum $(ru)^2$ stands for the energy of incompressible fluid in circular flow.

When the work done between fluid layers in the gap is zero, the mechanical energy could not be transmitted along the radial direction. Thus, the transfer of the angular momentum along the radial direction is interrupted. This is the reason why hydrodynamic turbulence cannot transport angular momentum effectively in astrophysical disks [36, 37].

## 3 Governing equations and numerical methods

The governing equations for the simulation of the Taylor-Couette flow between two concentric cylinders are the continuity equation and the filtered unsteady three-dimensional Navier-Stokes equations for incompressible fluid [38],

$$\nabla \cdot \mathbf{u} = 0. \tag{11}$$

$$\rho(\frac{\partial \mathbf{u}}{\partial t} + \mathbf{u} \cdot \nabla \mathbf{u}) = -\nabla p + \mu \nabla^2 \mathbf{u} + \mathbf{f}. \tag{12}$$

Here $\mathbf{u}$ is the velocity vector, $p$ is the static pressure, $\rho$ is the fluid density, $\mu$ is the dynamic viscosity and $\mathbf{f}$ is the gravitational force.

The wall boundary condition on the static cylinder and on the rotating cylinder is respectively as,

$$\mathbf{u} = 0. \quad \text{at outer cylinder} \tag{13}$$
$$\mathbf{u} = \mathbf{u}_1. \quad \text{at inner cylinder} \tag{14}$$

where $\mathbf{u}_1$ is the velocity at the inner cylinder. Periodical boundary condition is applied at the two ends of z direction of the computational domain.

For the large eddy simulation (LES), the Wall-Adapting Local Eddy Viscosity(WALE) model for small scale turbulence is employed [39]. The LES method has been employed to simulate the Taylor-Couette flow in several previous studies, and reliable simulation results have been obtained [40, 41, 32].

The numerical simulation is carried out by a finite volume method with structured polygon meshes in a co-located grid. The SIMPLER (Semi-Implicit Method for Pressure-Linked Equations Revised) algorithm is employed to couple the solutions of these equations. The convected term in the governing equations is discretized with the second-order upwinding scheme and the diffusion term is approximated with the second central-difference scheme.

The accuracy of the numerical simulation result has been validated with experiment data [32]. Grid independence is verified with five sets of meshes, and finally a grid of 4194304 nodes is employed. The numbers of grids in tangential, radial and axial direction, are 256, 64, and 256, respectively.

The resolution of the computation can be judged by the maximum of y+ of the first layer of grid near the wall, where $y+ = \sqrt{\tau_w/\rho}$, $\tau_w$ is the shear stress at the wall and $\rho$ is the density. In present study, the value of y+ of the first layer grid near the walls is 0.09, and the expansion rate toward leaving the wall is 1.1.

Table 1 Locations of seven different monitoring points in the radial direction

| Monitoring points | 1 | 2 | 3 | 4 | 5 | 6 | 7 |
|---|---|---|---|---|---|---|---|
| r(m) | 0.044 | 0.045 | 0.046 | 0.048 | 0.0497 | 0.051 | 0.052 |
| (r-$R_1$)/d | 0.0674 | 0.1797 | 0.2921 | 0.5169 | 0.7079 | 0.8539 | 0.9663 |

The original geometry for the computation is taken from the work [42]. The radius of the inner cylinder and the outer cylinder are 0.0434 m and 0.0523 m, respectively. Thus, the radius ratio is $R_1/R_2$=0.83. The ratio of the length of the annulus to the gap width is 46.6.

The axial position of the monitoring points at the plane z=0.131 (dimensionless position is z/L=0.5) is selected at the middle of the axial direction. If the selected monitoring position moves a little along the axial direction, the result is similar.

In the following, the components of the velocity in the tangential, radial, and axial directions are expressed by $u_\theta, u_r$, and $u_z$, respectively. During the simulation, six different locations along the radial direction were selected as monitoring points in the plane z=0.131. The monitoring points are shown in Table 1.

## 4. Inverse energy cascade in turbulent Taylor-Couette flows

Numerical simulations for the Taylor-Couette flow are performed using the Large Eddy Simulation (LES) technique at four different Reynolds numbers (Re=600, 1000, 1600, 2500). To study the internal flow mechanisms of energy transfer at varying Reynolds numbers, six monitoring points are selected along the radial direction within the gap, as listed in Table 1. The energy spectra at these points are calculated and analyzed. Figure 3 shows the turbulent kinetic energy versus the fluctuation frequency at six different locations along the radial direction under the specified Reynolds numbers. The high frequency corresponds to the larger wave number.

The definition of turbulent kinetic energy is as follow,

$$E_d = \frac{1}{2}\left(u'^2 + v'^2 + w'^2\right). \tag{15}$$

where $u'$, $v'$, and $w'$, is the fluctuation velocity in the tangential, radial, and axial directions, respectively.

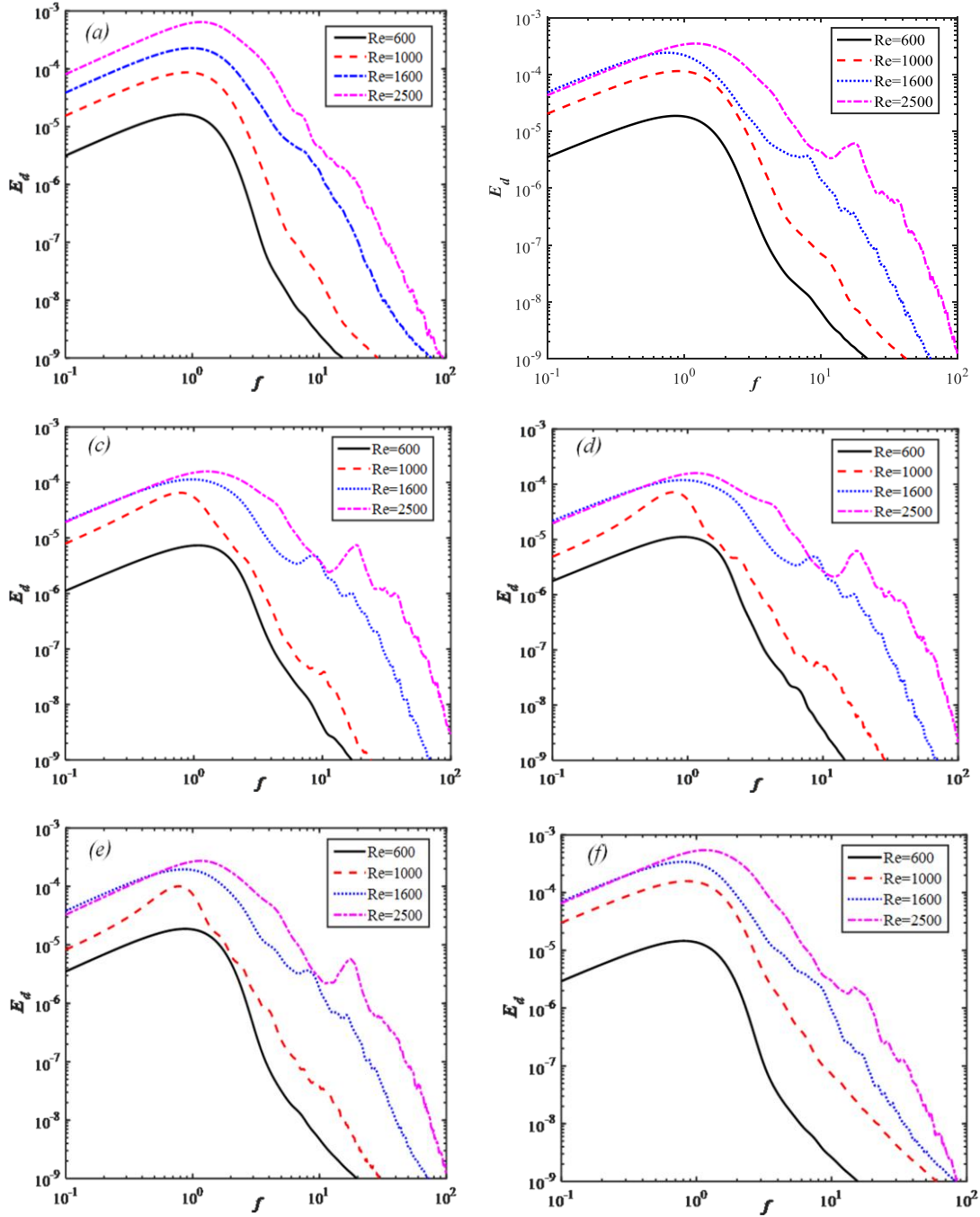

**Figure 3** Energy spectrum at six different locations along the radial direction at different Reynolds numbers, (a) to (f) corresponds to the monitoring points 1-6 in Table 1, respectively. The ordinate is the turbulent kinetic energy and the abscissa is the frequency of the velocity fluctuation.

From Figure 3, it can be seen that when the Reynolds numbers are Re=600 and Re=1000, the turbulent energy spectrum does not show any inverse cascade phenomenon at all points in the flow channel. When Re=1600, the inverse cascade phenomenon first appears in the core area of the flow channel. When Re=2500, the inverse cascade phenomenon extends from the core area of the channel to the wall area. It can also be seen that when the Reynolds number changes from 1600 to 2500, the peak turbulent kinetic energy of the inverse cascade phenomenon shifts from the low wave number region to the high wave number region. The phenomenon of peak turbulent kinetic energy indicates that within a certain range of the wave number, energy is concentrated in small-scale structures and no longer transmitted to the larger wave number. These phenomena are consistent with the inverse energy cascade phenomenon described in references [21, 29, 43-45], confirming the existence of inverse energy cascade in Taylor-Couette flow.

For Re=600, it can be seen that the energy spectrum curve shows a typical attenuation trend. All six energy spectrum curves at the monitoring location exhibit a monotonically decreasing pattern with increasing frequency, and no signs of inverse energy cascade are observed.

For Re=1000, Figure 3(d) shows a local small-scale peak in the frequency range of $2<f<3$ at monitoring point 4, indicating that an inverse energy cascade phenomenon has occurred at this location, while other monitoring points do not exhibit similar characteristics, suggesting that the inverse energy cascade phenomenon first occurs in the core region of the gap. This phenomenon may stem from the enhancement of vortex structures at specific scales, implying a local anomaly in the distribution of energy.

For Re=1600, significant small-scale peaks are observed at monitoring points 3, 4, and 5 in Fig.3(c), (d), and (e) within the frequency range of $6<f<9$, while the peaks at monitoring points 2 and 6 in Fig.3(b) and 3(f) are not pronounced. This result indicates that the core region of the gap where monitoring points 3, 4, and 5 are located is more susceptible to the inverse energy cascade phenomenon. Compared to Re=1000, the region exhibiting inverse energy cascade phenomenon has expanded in radial direction.

At higher Reynolds numbers (Re=2500), peaks are observed across a broader frequency range ($10<f<20$) at monitoring points 2, 3, 4, 5, and 6, with the peaks becoming more pronounced closer to the center of the gap. This indicates that the inverse energy cascade phenomenon occurs at higher frequency range, signifying that it is not limited to the core region of the gap but gradually spreads towards the inner and outer cylindrical walls. Compared to the core region of the gap, the extent of inverse energy cascades near the walls is relatively weaker.

In summary, the Taylor-Couette flow exhibits inverse energy cascade phenomenon under high Reynolds number conditions, initially mainly concentrated in the core region of the gap. As the Reynolds number increases, the range of inverse energy cascade phenomenon expands along the radial direction. Ultimately, this inverse energy cascade phenomenon covers the region near the inner and outer cylinders, although the level of inverse energy cascade in the wall region is relatively low. At the same time, as the Reynolds number increases, the frequency range of the inverse energy cascade shifts towards higher frequency.

**5. Physical mechanism of the inverse energy cascade in Taylor-Couette flow**

In the statistical theory of turbulence, the K41 theory due to Kolmogorov mainly explains the cascade of turbulent energy from larger to smaller scales and its dissipation due to viscosity. However, the K41 theory is not able to explain the phenomenon of inverse energy cascade. The energy gradient theory [30-32] demonstrated that the singularities of the Navier-Stokes (NS) equations lead to velocity discontinuities and the formation of spikes. The negative spikes occurring in temporal distribution of the streamwise velocity have been confirmed are the source and power of turbulent generation [30-32]. The occurrence of these spikes result in the cycle of turbulence burst, leading to the formation of large-scale coherent structures. It is shown that the energy of the large-scale vortices originates from the spikes of the NS equations, which transfer the energy from the averaged flow to the fluctuations of velocity components.

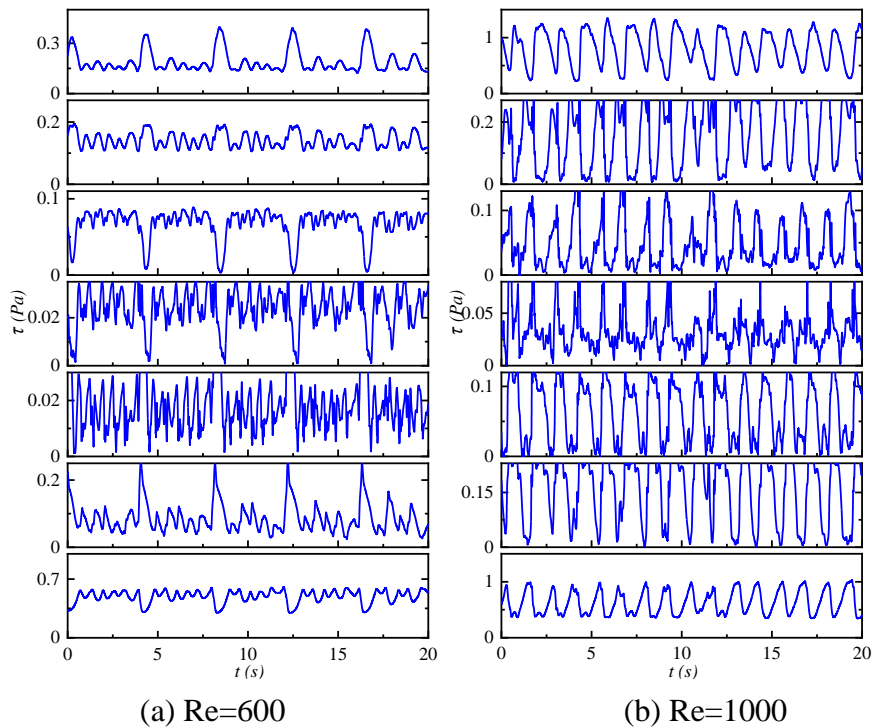

(a) Re=600  (b) Re=1000

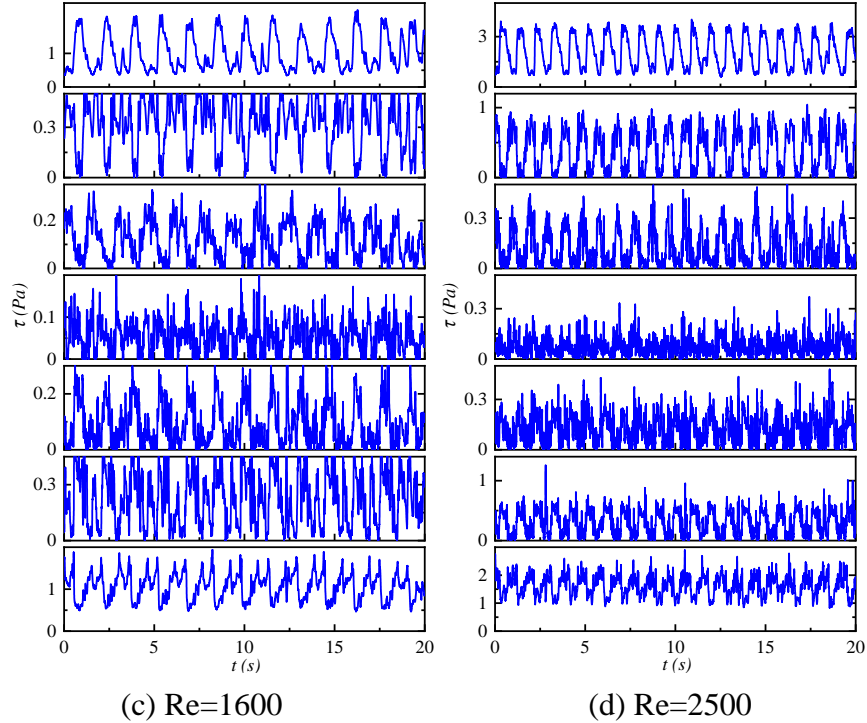

(c) Re=1600    (d) Re=2500

**Figure 4** Instantaneous distribution of shear stress along the radial direction at seven different monitoring points under various Reynolds numbers, with monitoring points 1 through 7 represented from bottom to top.

In Taylor-Couette flow, the evolution of the shear stress with time may result in the occurrence of τ=0, which leads to singularities in the Navier-Stokes (NS) equations and the energy gradient function K value. In this study, the instantaneous shear stress at different radial positions within the gap was monitored in detail.

Figure 4 shows the distribution of instantaneous shear stress at seven different monitoring points along the radial direction under various Reynolds numbers, representing monitoring points 1 through 7 from bottom to top. As shown in Figure 4 (a), when Re=600, the shear stress at a single monitoring point 4 begins to approach zero, but no case of τ=0 was observed. This indicates that the energy transfer between fluid layers can effectively occur through shear stress.

It is observed in Fig. 4(c) that the frequency of instantaneous shear stress τ=0 occurrences at monitoring points 2 and 6 is lower compared to those at monitoring points 3, 4, and 5. This discrepancy is directly related to the extent of energy aggregation, with monitoring points 2 and 6 showing significantly less energy aggregation compared to the latter. Correspondingly, in the energy spectrum curves presented in Fig.3(b) and Fig.3(f), the peaks are less pronounced compared to the distinct peaks recorded at other monitoring points in Fig.3(c), (d) and (e). This phenomenon indicates that events of shear stress τ=0 are positively correlated with the extent of energy aggregation, which in turn determines the shape of the energy spectrum curves.

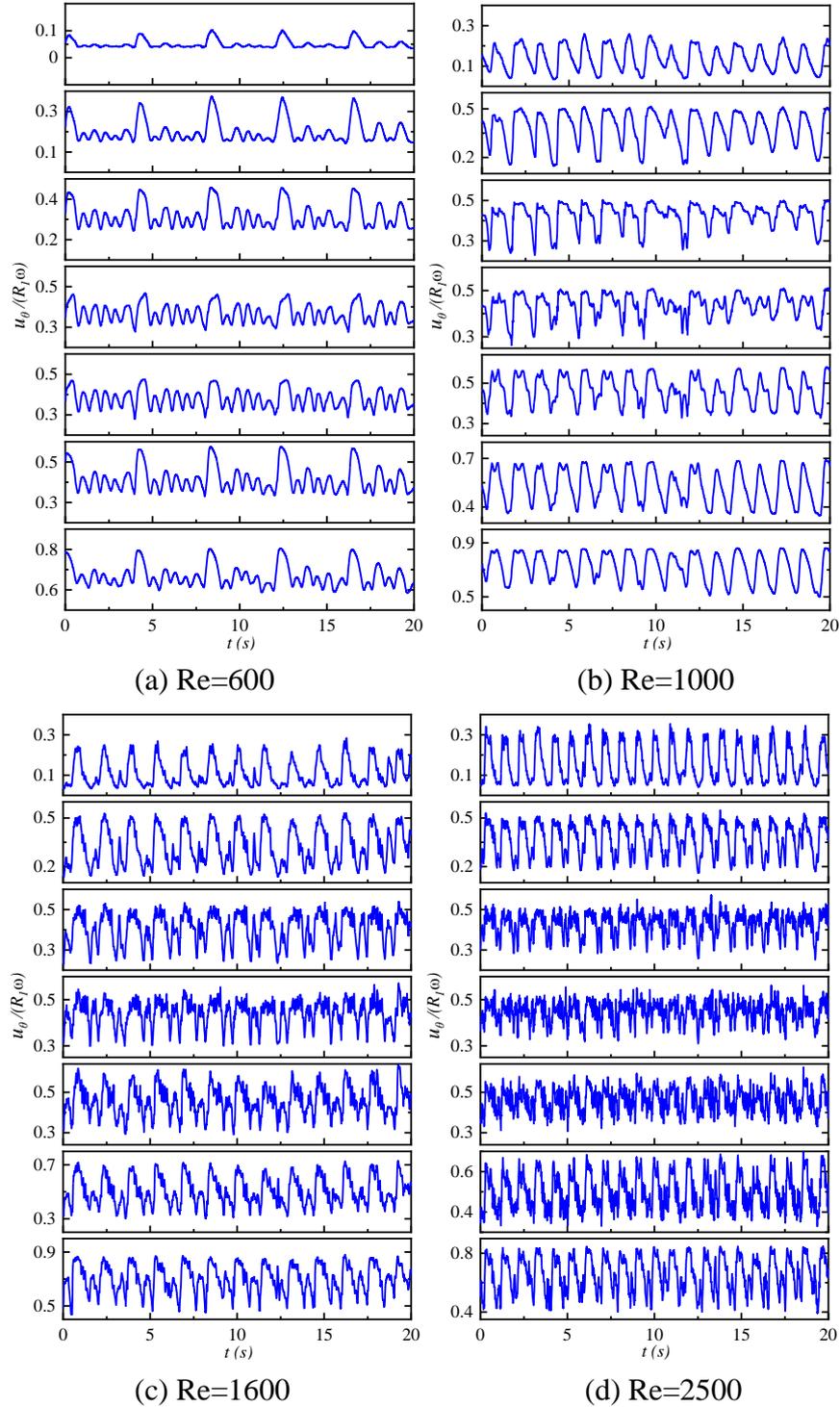

**Figure 5** Instantaneous distribution of tangential velocity along the radial direction at seven different monitoring points under various Reynolds numbers, with monitoring points 1 through 7 represented from bottom to top.

Figure 5 shows the instantaneous distribution of the tangential velocity along the radial direction at seven different monitoring points under various Reynolds numbers. It can be seen from Fig.5(a) that for Re=600, the instantaneous tangential velocity at each monitoring point rapidly decays and stabilizes near a certain value after exhibiting peaks, indicating that the energy of the large-scale vortices produced by velocity peaks can be smoothly transferred to small-scale vortices under the action of shear stress. Therefore, under this condition, the energy can be transferred from large-scale vortices to small-scale vortices, which is known as a direct energy cascade. This is illustrated by the black solid line in Figure 3, where the turbulent energy curves at all monitoring points exhibit a typical decaying trend.

It can be observed from Fig.5(b) that for Re=1000, negative spikes in instantaneous tangential velocity are generated, in particular, after t=10 seconds. After that, the tangential velocity exhibits low-frequency, low-amplitude oscillations and gradually decays, suggesting that the energy transfer in the fluid near the mid-gap is impeded, preventing effective transmission from large-scale vortices to smaller ones. This results in the accumulation of turbulent energy near the middle of the gap, consistent with the small-scale peak feature (f=7~10) shown by the red dashed line in Fig.3(d). In addition, it can be observed from Fig.4(b) that for Re=1000, instances of shear stress $\tau=0$ are observed at the monitoring point 4, marking a singularity in the Navier-Stokes (NS) equations at this location, which initiates instabilities in the flow. However, apart from the monitoring point 4, no other monitoring points along the radial direction display a reduction of shear stress $\tau$ to zero (Fig.4(b)). This ensures smooth energy transfer between adjacent fluid layers and allowing the energy of large-scale vortices to be efficiently transformed to small-scale vortices, with their energy spectrum curves exhibiting a typical decay trend.

It can be observed from Fig.5(c) that for Re=1600, in the instantaneous tangential velocity distribution, the tangential velocity at the monitoring points 3, 4, and 5 shows periodic spikes, accompanied by a slow decay after high-frequency oscillations. This indicates a reduced rate of energy dissipation of large eddies in this area, leading to local energy accumulation. In contrast, the instantaneous tangential velocity at other monitoring points decays more rapidly to a lower level after reaching their peak and oscillates around this velocity value, achieving effective energy attenuation and dissipation.

This feature of the tangential velocity can be further analyzed from Fig.4(c) at Re=1600. It can be observed that the fraction of time with the shear stress $\tau=0$ at monitoring points 3, 4, and 5 have increased, while the shear stress at monitoring points 2 and 6 exhibit similar phenomenon at certain time fractions, indicating that flow instability (shear stress $\tau=0$) is gradually expanding from the mid-gap to the positions at these monitoring points. Due to the more frequent occurrences of shear stress $\tau=0$ at monitoring points 3, 4, and 5, compared to the case of Re=1000, energy transfer in the core region of the gap for Re=1600 becomes more impeded.

The shear stress at Re=2500 in Fig.4(d) shows more time fraction with $\tau=0$ than those in the case of Re=1600, at monitoring points 2, 3, 4, and 5. The instantaneous tangential velocity at these monitoring points all exhibits periodic spike. In the distribution of instantaneous velocity, the frequency of small-amplitude and high-frequency oscillations at higher speeds becomes even higher, as shown in Fig.5(d). This observation suggests that the locations corresponding to these monitoring points are now in a state of flow instability. This phenomenon indicates that the regions where the energy transfer is hindered are more widespread at these positions, and the efficiency of turbulent energy transfer has decreased significantly.

Meanwhile, in Fig.5(d), it is observed that the frequency of the high-frequency oscillations with small amplitude at monitoring points 2 and 6 are lower compared to those at monitoring points 3, 4, and 5. This disparity is reflected in the analysis of the energy spectra: the frequency corresponding to the small-scale peak in the energy spectrum curves under the condition of Re=2500 in Fig.3(b) and 3(f) is lower than that in Fig.3(c), 3(d), and 3(e). This phenomenon reveals a positive correlation between the frequency position of the peaks on the energy spectrum curves and the frequency of the high-frequency oscillations with small amplitude.

In turbulent Taylor-Couette flow, concerning the formation of the inverse energy cascade and the small-scale peaks in the energy spectrum distribution, we infer that the formation of small-scale peaks in the energy spectrum distribution is closely related to the singularity of the Navier-Stokes (NS) equations, which results in shear stress $\tau=0$. Within regions of the flow gap where the shear stress $\tau=0$ occurs, the outward (radial) transfer of turbulent energy is suppressed, leading to an energy accumulation in small-scale vortices within specific frequency ranges, and ultimately forming peaks in the energy spectrum distribution. The more singularities there are in the Navier-Stokes equations, the more extensive the distribution of regions with shear stress $\tau=0$ in the Taylor-Couette flow. This widespread distribution exacerbates flow instability, making the peak features in the energy spectrum more pronounced and the inverse energy cascade phenomenon more evident.

As discussed above, the mechanism of the inverse energy cascade is the instantaneous zero shear stress in the core area of the cylinder gap. It is summarized as follow for clarity. It is observed from Fig.4 that when Re=600, there is no occurrence of $\tau=0$, and the Navier-Stokes equation does not exhibit singularity. There are no small peaks in the corresponding energy spectrum (Fig.3). When Re=1000, the simulation result indicates a situation where $\tau=0$ at monitoring point 4, which exhibits singularities of the Navier-Stokes equation. The red dashed line in Fig.3(d) shows local small-scale peaks in the frequency range of $2<f<3$, indicating the occurrence of inverse energy cascade at this location. When Re=2500, the region with zero shear stress expands radially compared to the cases of Re=1600 (Fig.5). The phenomenon of small-scale peaks appearing in the core area also extends radially. This means that zero shear stress is the reason for the occurrence of inverse energy cascade.

Figure 6 shows the streamlines at r-z plane at five times in a period for the flow for various Reynolds numbers. The contours shown in the figure are the tangential velocity. The upper boundary is the outer cylinder and the bottom boundary is the inner cylinder. As mentioned previously, it can be observed from Fig.5 that due to the obstruction of turbulent energy transfer in the core region of the gap, the high-frequency fluctuations with small amplitude are observed in the instantaneous tangential velocity distribution, shown as the monitoring points 4 and 5 in Fig.5(c) for Re=1600 and Fig.5(d) for Re=2500. These features are also observed in Fig.6(c) and Fig.6(d), where multiple small-scale Taylor vortices are detected within some of the large-scale turbulent Taylor vortices. These findings are consistent with the research results of Froitzheim and Lohse [29]. It is precisely the aggregation of these small vortices with high turbulent energy in the core region of the gap that leads to the appearance of small spikes in a specific frequency range in the energy spectrum, as indicated in Figures 3(c), (d), and (e) for Re=1600 and Re=2500. The frequency range in which these small-scale peaks appear is consistent with the frequency of velocity fluctuations with small-amplitude.

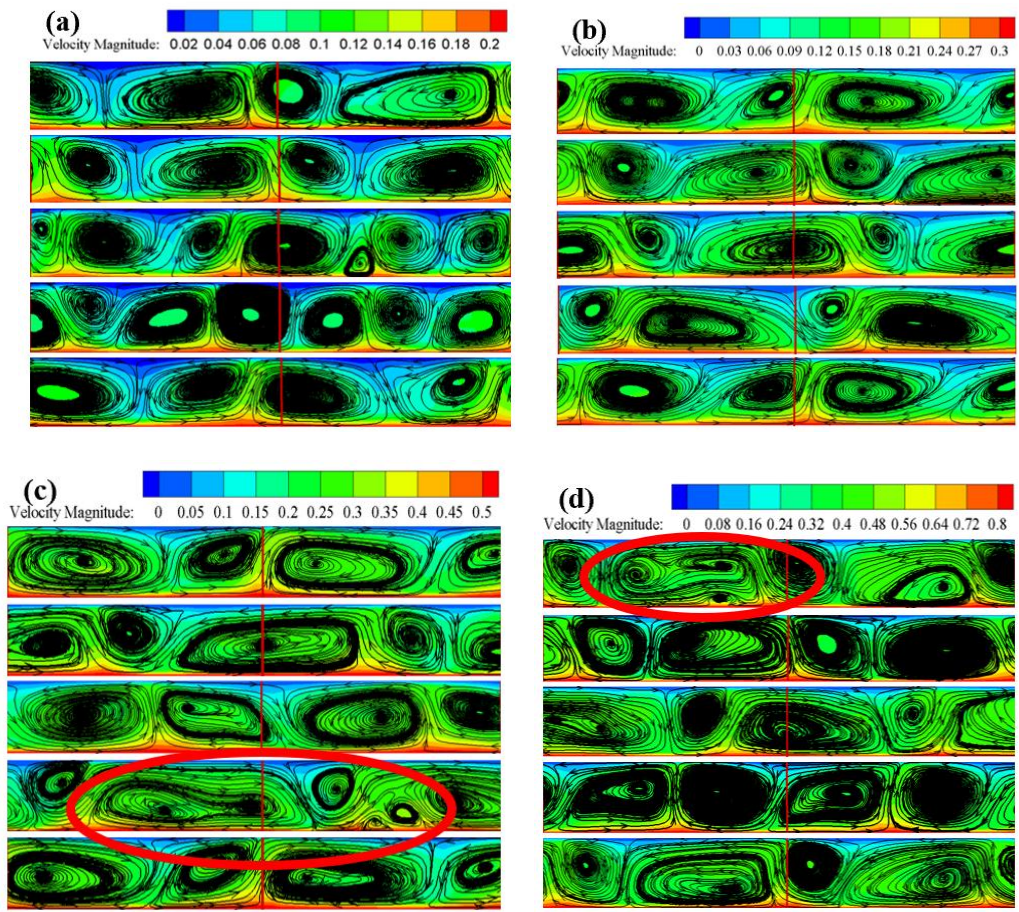

**Figure 6** Streamlines at r-z plane at five time moments in a period for the flow at different Reynolds numbers, (a)Re=600, (b)Re=1000, (c)Re=1600, (d)Re=2500. The contours shown in the figure are the tangential velocity. The upper boundary is the outer cylinder and the bottom boundary is the inner cylinder.

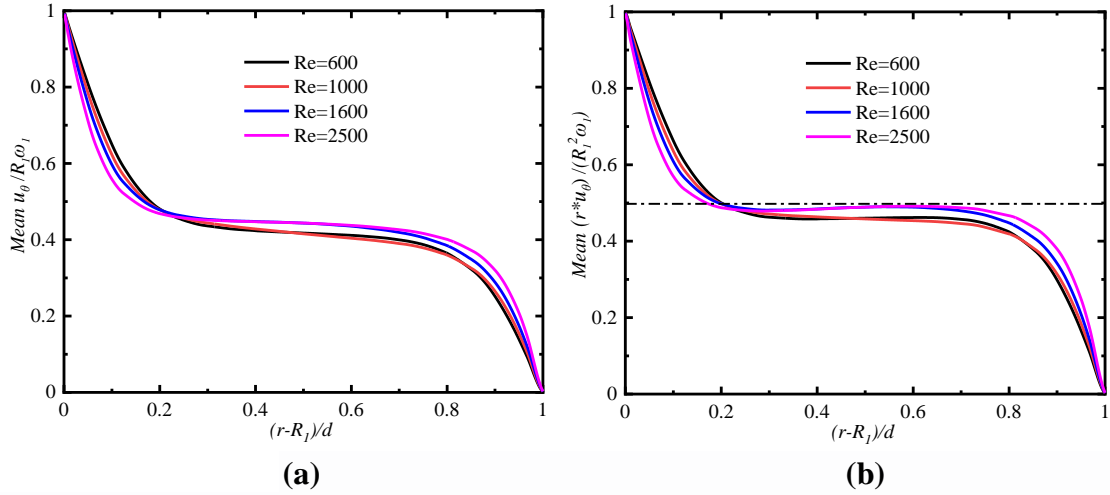

**Figure 7** Radial distributions of (a) mean tangential velocity and (b) mean angular momentum at z=0.131 m under varying Reynolds numbers.

Moreover, it is observed that the velocity fluctuations with small-amplitude significantly excite smaller-scale vortices carrying turbulent energy within the larger-scale vortices, as shown in Fig.6(d). This excitation of smaller-scale vortices further corroborates the complex mechanisms of turbulent energy transformation within large-scale structures.

Figure 7 illustrates the radial distribution of the mean tangential velocity and the mean angular momentum at z=0.131m under four different Reynolds numbers. It is observed from the radial distribution of the mean tangential velocity that the tangential velocity within the core region of the gap exhibits a flat profile with a very small velocity gradient, which is close to zero. As the Reynolds number increases, the flat part of the tangential velocity in the core region of the gap widens and extends closer to the two cylindrical walls.

It is also observed that the mean velocity distribution in Fig.7 shows a non-monotonic Reynolds number dependence near the inner cylinder, which seems abnormal. This phenomenon may be explained as follow. Near both the cylinder walls, the velocity gradient increases with the Re. Thus, across the gap width from the inner cylinder to the outer cylinder, there must be intersection of these curves from low to high Re. After enlargement view of the velocity near the inner cylinder walls, it is clear that the velocity gradient near the inner cylinder is really increasing with Re. Therefore, these results of mean velocity distributions are reasonable.

For the radial distribution of the mean angular momentum in Fig.7, at Re=600 and Re=1000, the variation of mean angular momentum in the core region of the gap is relatively more gradual compared to the regions near the two cylindrical walls. Specifically, at Re=600, the mean angular momentum initially increases and then decreases in the core region of the gap; whereas at Re=1000, the mean angular momentum exhibits a characteristic of gradual decrease in the core region of the gap. However, when the Reynolds number increases to Re=1600 and Re=2500, the mean angular momentum in the core region of the gap remains nearly constant, particularly at Re=2500, where the region over which the mean angular momentum stays constant significantly expands. Since the mean angular momentum remains constant in the core region of the gap, this implies that the energy cannot be effectively transferred within this area. This explains why the phenomenon of inverse energy cascade first emerges in the core region of the gap and then extends towards the regions near the cylindrical walls as the Reynolds number increases.

In contrast, the radial gradients of the total mechanical energy are considerably larger in the regions adjacent to the cylindrical walls, which facilitate the smooth transfer of energy. Consequently, these regions do not exhibit pronounced phenomena of energy accumulation. This difference explains why the wall regions demonstrate weaker inverse energy cascade effects compared to the core region.

## 6. Conclusions

In this study, LES work is performed to study the inverse energy cascade occurring in the gap of the Taylor-Couette flow between two coaxial cylinders. The outer cylinder is fixed and the inner cylinder is rotating. Simulations for four Reynolds numbers from 600 to 2500 have been performed, and the physical mechanism of the inverse energy cascade is investigated.

The inverse energy cascade was found to occur in rotating flow and it has a hindering effect on energy transfer in the gap in previous studies. The energy transfer and heat conduction in rotating flow have a significant impact on the flow characteristics. For a long time, the mechanism of the inverse energy cascade has not been clarified. In this study, it is discovered that the mechanism of the inverse energy cascade is caused by the instantaneous zero shear stress. Therefore, the principle of the inverse energy cascade can be extended to other flow applications to achieve the goal of flow control. For example, it has been found in astrophysics that it is difficult to transfer energy in rotating systems. There may be due to effect of the inverse energy cascade. Inverse energy cascade has also been found to exist in the flow of MHD, where the coupling between electromagnetic force and flow leads to complex flow. Therefore, understanding the mechanism of the inverse energy cascade can enhance the utilizations of the internal fluid flow, energy transfer, and heat conduction in rotating flow and MHD systems.

The following conclusions can be drawn from present study.

(1) As the Reynolds number increases, the Taylor-Couette flow exhibits inverse energy cascade phenomenon, which is initially concentrated in the core region of the gap. As the Reynolds number further increases, this phenomenon extends to the inner and outer cylindrical walls, but compared to the core region, the degree of inverse energy cascade in the wall regions is relatively weaker.

(2) This study provides a comprehensive analysis of the inverse energy cascade phenomenon in turbulent Taylor-Couette flows, revealing the intricate relationship between the singularity of the Navier-Stokes equations and the dynamics of energy transfer. Numerical simulation results demonstrate that the formation of the small-scale peak in the energy spectrum distribution is closely related to the instantaneous shear stress $\tau=0$ induced by the singularity of the Navier-Stokes equations.

(3) Compared to the core region of the gap, the radial gradients of the total mechanical energy in the regions adjacent to the two cylindrical walls are significantly larger, facilitating smoother energy transfer in these areas. These regions do not exhibit obvious energy accumulation phenomenon, hence the intensity of the inverse energy cascade near the walls is relatively weaker. Additionally, it is observed that with the increase of the Reynolds number, the frequency range in which the inverse energy cascade occurs shifts to higher frequency.

(4) High-frequency fluctuations with small amplitude are observed in the instantaneous tangential velocity distribution. This is caused by the obstruction of turbulent energy transfer in the core region of the gap. This phenomenon leads to the excitation of small vortices with turbulent energy occurring within the large vortices, whose scales are smaller than the large-scale vortices but much larger than the dissipation scale ones. Additionally, it is found that the frequency of the appearance of small-scale peak in the energy spectrum is related to the frequency of small-amplitude high-frequency oscillations.

(5) Within the gap of the Taylor-Couette flows, the distribution of the mean angular momentum determines the energy transfer along the radial direction. When the gradient of the mean angular momentum in the radial direction is reduced, the rate of the energy transfer along the radial direction decreases. When the mean angular momentum becomes constant, the energy transfer along the radial direction is impeded. The phenomenon of inverse energy cascade is closely related to this factor.

(6) It is found that there is no transfer of turbulent energy from small vortices to large vortices in the inverse energy cascade, which is different from some old concept, but the accumulation of small-scale vortices with high turbulent energy within large-scale vortices at special flow environment.

**Acknowledgement**

We thank the organization of the The Ninth International Symposium on Physics of Fluids of the State Key Laboratory of Turbulence and Complex Systems of Peking University, National University of Singapore, Nanjing University of Aeronautics and Astronautics, Taiyuan University of Technology, and AIP Publishing in Datong, China held over 12-15 June of 2024 for creating the platform at which, and for bringing together the audience to which, this work was first presented. The authors thank Prof. F. J. Wang for the helpful discussions. This work was partialy supported by the research fund of Hangzhou Dianzi University (KYS015623111).

**Compliance with Ethical Standards**
   **Conflict of Interest:** The authors declare that they have no conflict of interest.
   **Funding:** There is no funding source.
   **Ethical approval:** This article does not contain any studies with human participants or animals performed by any of the authors.
   **Informed consent:** Informed consent was obtained from all individual participants included in the study.

**References**
1. U. Frisch, Turbulence: The Legacy of A. N. Kolmogorov (Cambridge University Press, Cambridge, UK, 1995).
2. L. F. Richardson, Weather Prediction by Numerical Process (Cambridge University Press, Cambridge, UK, 1922).
3. A. N. Kolmogorov, "The local structure of turbulence in incompressible viscous fluid for very large Reynolds numbers," Cr Acad. Sci. URSS, 30: 301-305(1941).
4. R. H. Kraichnan, "Inertial ranges in two-dimensional turbulence," Physics of Fluids, 10(7): 1417-1423(1967).
5. G. Boffetta, S. Musacchio, "Evidence forthedoublecascade scenario in two-dimensional turbulence," Physical Review E, 82(1): 016307(2010).
6. G. Boffetta, R. E. Ecke, "Two-dimensional turbulence," Annual Review of Fluid Mechanics, 44(1), 427-451(2012).
7. J. Paret, P. Tabeling, "Intermittency in the two-dimensional inverse cascadeof energy: Experimental Observations," Physics of Fluids, 10(12): 3126-3136(1998).
8. K. Schneider, M. Farge . "On Decaying Two-Dimensional Turbulence in a Circular Container." Physical Review Letters, 95(24), 244502(2005).https://doi.org/10.1103/PhysRevLett.95.244502
9. A. W. Baggaley, C. F. Barenghi, Y. A. Sergeev. "Three-dimensional inverse energy transfer induced by vortex reconnections." Physical Review E, 89, 013002(2014). https://doi.org/10.1103/PhysRevE.89.013002
10. P. Bartello. "Self-similarity of decaying two-dimensional turbulence." Journal of Fluid Mechanics, 326, 357-372(1996).https://doi.org/10.1017/S002211209600835X
11. B. H. Burgess, D. G. Dritschel, R. K. Scott. "Vortex scaling ranges in two-dimensional turbulence." Physics of Fluids 29(11), 111104(2017). https://doi.org/10.1063/1.4993144
12. S. Chen, R. Ecke, G. Eyink, M. Rivera, M. Wan, Z. Xiao, "Physical mechanism of the two-dimensional inverse energy cascade," Phys. Rev. Lett. 96:084502(2006).
13. Z. Xiao, X. Wang, S. Chen, G. Eyink, "Physical mechanism of the inverse energy cascade of two-dimensional turbulence: a numerical investigation," J. Fluid Mech. 619:1-44(2009).
14. R. H. Kraichnan, R. Panda, "Depression of nonlinearity in decaying isotropic turbulence," Physics of Fluids, 31(31):2395-2397(1988).
15. Q. Chen, S. Chen, G. L. Eyink, D. D. Holm, "Intermittency in the joint cascade of energy and helicity," Physical Review Letters, 90(21), 214503(2003).
16. Q. Chen, S. Chen, G. L. Eyink. "The joint cascade of energy and helicity in three-dimensional turbulence", Physics of Fluids, 15, 361-374(2003). https://doi.org/10.1063/1.1533070
17. R. Hu, X. Li, C. Yu, "Transfers of energy and helicity in helical rotating turbulence,"
Journal of Fluid Mechanics, 946: A19(2022).
18. M. I. Cheikh, J. Chen, M. Wei, "Small-scale energy cascade in homogeneous

isotropic turbulence," Physical Review Fluids, 4(10) (2019).
19. X. Wit, M. D. Fruchart, M. Khain, T. Toschi, F. Vitelli, "Pattern formation by turbulent cascades," Nature, 627, 515-521(2024).
20. S. Galtier, Introduction to Modern Magnetohydrodynamics (Cambridge University Press, Cambridge, UK, 2016).
21. L. Biferale, S. Musacchio, F. Toschi, "Inverse energy cascade in three-dimensional isotropic turbulence," Physical Review Letters, 108(16), 164501(2012).
22. H. Xia, D. Byrne, G. Falkovich, M. Shats, "Upscale energy transfer in thick turbulent fluid layers," Nature Physics, 7(5), 321-324(2011).
23. J. Słomka, J. Dunkel, "Spontaneous mirror-symmetry breaking induces inverse energy cascade in 3D active fluids," Proceedings of the National Academy of Sciences of the United States of America, 114(9), 2119-2124(2017).
24. G. S. Lewis, H. L. Swinney, "Velocity structure functions, scaling and transitions in high Reynolds number Couette-Taylor flow," Phys. Rev. E, 59, 5457-5467(1999).
25. W. Polifke, L. Shtilman. "The dynamics of helical decaying turbulence." Physics of Fluids A Fluid Dynamics, 1, 2025-2033(1989). https://doi.org/10.1063/1.857476
26. S. Dong, "Direct numerical simulation of turbulent Taylor-Couette flow," J. Fluid Mech. 587, 373-393 (2007).
27. R. VanHout, and J. Katz, "Measurements of mean flow and turbulence characteristics inhigh-Reynolds number counter-rotating Taylor–Couette flow," Phys. Fluids 23, 105102(2011).
28. S. G. Huismans, D. Lohse, C. Sun, "Statistics of turbulent fluctuations in counter-rotating Taylor-Couette flows," Phys. Rev. E 88, 063001(2013).
29. A. Froitzheim, R. Ezeta, S. G. Huisman, S. Merbold, C. Sun, D. Lohse, C. Egbers, "Statistics, plumes and azimuthally travelling waves in ultimate Taylor-Couette turbulent vortices," J. Fluid Mech., 876, 733-765(2019).
30. H.-S. Dou. "Singularity of Navier-Stokes equations leading to turbulence," Adv. Appl. Math. Mech., 13(3), 527-553(2021).
31. H.-S. Dou, "No existence and smoothness of solution of the Navier-Stokes equation," Entropy, 24, 339(2022).
32. H.-S. Dou, Origin of Turbulence: Energy Gradient Theory (Springer, Singapore,2022).
33. H.-S. Dou, B. C. Khoo, K. S. Yeo, "Energy loss distribution in the plane Couette flow and the Taylor-Couette flow between concentric rotating cylinders," Inter. J. Therm. Sci., 46, 262-275(2007).
34. H.-S. Dou, B. C. Khoo, K. S. Yeo, "Instability of Taylor-Couette flow between concentric rotating cylinders," Inter. J. Therm. Sci., 47, 1422-1435(2008).
35. H.-S. Dou, "Mechanism of flow instability and transition to turbulence," Inter. J. Non-Linear Mech., 41 (4), 512-517(2006).
36. H. Ji, S. A. Balbus, "Angular momentum transport in astrophysics and in the lab," Phys. Today, 66(8), 27-33(2013).
37. H. Ji, M. Burin, E. Schartman, J. Goodman, "Hydrodynamic turbulence cannot

transport angular momentum effectively in astrophysical disks," Nature, 444, 343–346(2006).
38. F. M. White, Viscous Fluid Flow, 2nd Ed., (McGraw-Hill, New York, 1991).
39. H. K. Versteeg, W. Malalasekera, An Introduction to Computational Fluid Dynamics-The finite volume method, 2nd Edition (Pearson Education, England, 2007).
40. S. Poncet, S. Viazzo, R. Oguic, "Large eddy simulations of Taylor-Couette-Poiseuille flows in a narrow-gap system," Physics of Fluids, 26, 105108(2014).
41. M. A. Razzak, B. C. Khoo, K. B. Lua, "Numerical study of Taylor-Couette flow with longitudinal corrugated surface," Phys. Fluids, 32, 053606(2020).
42. S. T. Wereley, R. M. Lueptow, "Velocity field for Taylor–Couette flow with an axial flow," Phys. Fluids, 11(12), 3637-3649(1999).
43. G. Sahoo, L. Biferale, A. Alexakis, "Transition from Direct to Inverse Cascade in Three-Dimensional Turbulence," Phys. Rev. Lett., 118:164501(2017).
44. T. Pestana, S. Hickel, "Regime Transition in the Energy Cascade of Rotating Turbulence," Phys. Rev. E 99, 053103(2018).
45. J. I. Polanco, G. Krstulovic, "Counterflow-induced inverse energy cascade in three-dimensional superfluid turbulence," Phys. Rev. Lett., 125: 254504(2020).